\documentclass[12pt]{iopart}

\usepackage{iopams}
\usepackage{latexsym}
\usepackage{amssymb}
\usepackage[dvips]{graphicx}
\usepackage{color}
\begin{document}

\title[Mixing of Frenkel and charge-transfer excitons]{Mixing of Frenkel and charge-transfer excitons in quasi-one-dimensional one-component molecular crystals}

\author{I J Lalov and I Zhelyazkov}

\address{
Faculty of Physics, Sofia University, BG-1164 Sofia, Bulgaria}
\ead{izh@phys.uni-sofia.bg}

\begin{abstract}
In molecular crystals like MePTCDI and PTCDA the molecule are regularly arranged creating quasi-one-dimensional molecular stacks.  The intermolecular distance in a stack (about $3.3$ \AA) is comparable with the electron--hole distance in the excited molecule.  The mixing between Frenkel excitons (FEs) and charge-transfer excitons (CTEs) is very essential for the excitonic and vibronic spectra of both crystals.  In this paper, we make simulations of the linear absorption spectra of the abovementioned crystals.  The basic Hamiltonian describes the FE--CTEs mixing in the molecular stack (point group $C_i$) caused by two transfer mechanisms, notably of the electron and the hole on the neighbor molecules.  The vibronic spectra consist of mixed excitons and one vibrational mode of an intramolecular vibration linearly coupled with FE and CTEs.  Using the vibronic approach, we calculate the linear optical susceptibility in the excitonic and one-phonon vibronic regions of the molecular stack, as well as of a crystal which contains two types of nonequivalent stacks.  We put the excitonic and vibrational parameters for the crystals of PTCDA and MePTCDI fitted in the previous studies.  We analyze the general structure and some important features of the lineshape of the linear absorption spectra in the spectral region of $15\,000$--$23\,000$ cm$^{-1}$.  We vary the values of the excitonic linewidth and the parameters of the linear exciton--phonon coupling and look for the values which reproduce the absorption lineshape similar to the absorption spectra of the investigated crystals.  Our study exhibits the necessity of introducing the FE--CTEs mixing in the interpretation of the linear absorption spectra, especially of the MePTCDI crystal.
\end{abstract}

\pacs{71.35.Aa, 73.20.Mf, 78.40.Me}
\submitto{\NJP}
\maketitle

\section{Introduction}
\label{sec:intro}
The mixing of Frenkel excitons (FEs) and charge-transfer excitons (CTEs) has been studied both experimentally and theoretically in many one-component molecular crystals, e.g., polyacenes \cite{sebastian81,sebastian83,siebrand83,petelenz96}, perilene derivatives \cite{henessy99,hoffmann00,hoffmann02}, fullerenes \cite{jeglinski92,pac98}, and other.  The excitonic and vibronic spectra of quasi-one-dimensional crystals like 3,4,9,10-perylenetetracarboxylic dianhydride (PTCDA) and $N,N^{\prime}$-dimethylperylene-3,4,9,10-perylenetetracarboximide (MePTCDI) have been treated in \cite{henessy99,hoffmann00,hoffmann02} and \cite{schmidt02,lalov06a,lalov07a,lalov06b,lalov06c}.  The short molecular distance in quasi-one-dimensional stacks causes a strong FE--CTEs mixing and mixing of their vibronic satellites as well, especially in the absorption spectra in the spectral region of $2$--$3$ eV ($15\,000$--$23\,000$ cm$^{-1}$).

In the present paper, we calculate the linear absorption spectra of one-component MePTCDI and PTCDA crystals applying the vibronic approach  developed in our previous papers \cite{lalov06b,lalov07b}.  It allows complex calculations of the absorption in pure excitonic, one-phonon vibronic, two-phonon vibronic, etc.\ spectra.  In paper \cite{lalov08} the vibronic approach is the tool of studying the FE--CTEs mixing in a two-component stack of alternatively arranged donor--acceptor (DA)-molecules.  While the FE--CTEs mixing in \cite{lalov08} is a probable but hypothetical model, in the present study we turn to the real absorption spectra and use the parameters of the FE--CTEs mixing from Refs.~\cite{henessy99,hoffmann00,schmidt02}.

Two general differences exhibit the models' comparing of the FE--CTEs mixing in one- and two-component molecular stacks \cite{lalov08}, notably: (i) The intermolecular transfer both of electrons and holes must be a feature of the model for the case of one-component stack whereas only one transfer mechanism is sufficient in a DA-molecular stack.  In the last case the Frenkel exciton represents a collectivized electronic excitation of a donor or an acceptor and the strongest transfer on the closest neighbor would be either of a hole or correspondingly of an electron.  Obviously in an one-component stack both types of transfer on the neighbors could be of the same probability.  The two-step processes of successive transfer of the electron and the hole on the same molecule ensure the transfer of a FE even in the case of its relatively weak direct intermolecular transfer \cite{schmidt02}.  In this way, our present study extends the calculations in \cite{lalov06b} where only one transfer mechanism has been considered. (ii) In the most widely studied DA-crystal, anthracene-PMDA the excitonic absorption lines are narrow \cite{haarer75,brillante80,weiser04} and many details in the vibronic spectra can be seen.  In PTCDA and MePTCDI crystals the absorption lines are two order of magnitude wider.  Thus we pay attention to the general structure of the excitonic and one-phonon vibronic spectra supposing a width of the excitonic lines of $300$--$500$ cm$^{-1}$ (not $2$--$10$ cm$^{-1}$ which is the absorption width of the anthracene-PMDA).

In the next section of the paper we involve the initial Hamiltonian in the case of a FE--CTEs mixing.  The Hamiltonian contains one mode of intramolecular vibration linearly coupled to the FE and CTEs.  In section 3 the linear optical susceptibility has been calculated in the excitonic and one-phonon vibronic spectra.  In section 4 the linear absorption has been modelled using the excitonic and vibrational parameters of the PTCDA and MePTCDI crystals fitted in Refs.~\cite{hoffmann00,hoffmann02,schmidt02}.  Section 5 contains some conclusions.

\section{Hamiltonian for the case of a FE--CTEs--phonon coupling}
We consider the excitonic and vibronic excitations in a linear molecular stack of $N$ identical molecules which are regularly arranged at a distance $d$ each other.  The point group of symmetry of the stack is $C_i$ as is for the molecular stacks of PTCDA and MePTCDI \cite{hoffmann00}.  The origin of the Frenkel exciton is a non-degenerate molecular electronic excitation with excitation energy $E_{\rm F}$ and $L$ being the transfer integral between neighboring molecules.  We denote by $B_n$ ($B_n^{+}$) the annihilation (creation) operator of the electronic excitation on molecule $n$ and get the following FE-part of the Hamiltonian:
\begin{equation}
\label{eq:fehamilt}
    H_{\rm FE} = \sum_n E_{\rm F} B_n^{+}B_n + \sum_{n,n^{\prime}} L \left( \delta_{n^{\prime},n+1} + \delta_{n^{\prime},n-1} \right) B_{n^{\prime}}^{+}B_n.
\end{equation}
As usually, we consider two CTEs of equal excitation energy $E_{\rm c}$ and $C_{n,1}$ is the annihilation operator CTE, $\sigma =1$, with hole located on the site $n$ and electron on the site $n + 1$, whereas the electron of the second CTE, $\sigma =2$, ($C_{n,2}$) is located at molecule $n - 1$.  We neglect the transfer of CTEs as a whole and the mutual coupling of the two CTEs since those processes can be realized through the transfer of the electron or hole at distance $2d$ which is less probable than the FE--CTEs mixing caused by the electron (hole) transfer at the neighbor molecule (see Refs.~\cite{lalov08,brillante80}).  We obtain the following CTEs-part of the Hamiltonian:
\begin{equation}
\label{eq:cteshamilt}
    \hat{H}_{\rm CT} = \sum_{n,\sigma = 1,2} E_{\rm c}C_{n \sigma}^{+}C_{n \sigma}
\end{equation}
and suppose the following operator for the FE--CTEs mixing:
\begin{eqnarray}
\label{eq:fcoperator}
    \hat{H}_{\rm FC} = \sum_n \left[ \varepsilon_{\rm e1}B_n^{+}C_{n,1} + \varepsilon_{\rm e2}B_n^{+}C_{n,2} \right. \nonumber \\
    \left.
    {}+ \varepsilon_{\rm h1}B_n^{+}C_{n-1,1} + \varepsilon_{\rm h2}B_n^{+}C_{n+1,2} + \mbox{h.c.} \right],
\end{eqnarray}
where $\varepsilon_{\rm e1}$ and $\varepsilon_{\rm e2}$ are the transfer integrals of the electron from molecule $n$ to molecules $n+1$ and $n-1$ correspondingly, and $\varepsilon_{\rm h1}$, $\varepsilon_{\rm h2}$ denote the transfer integrals of the hole from molecule $n$ to molecules $n+1$ and $n-1$.  Certainly the model with four transfer integrals is more complicated than the model in Refs.~\cite{hoffmann00,hoffmann02} with two mixing parameters only.  But our model is more realistic because we take into account the inclination of the flat molecules of PTCDA and MePTCDI relatively to the stack axis.

One intramolecular mode is only supposed to be coupled with the FE and CTEs and the phonon part of the Hamiltonian is
\begin{equation}
\label{eq:phonhamilt}
    \hat{H}_{\rm ph} = \sum_n \hbar \omega_0 a_n^{+} a_n,
\end{equation}
where $\omega_0$ is the vibrational frequency and $a_n$ is the annihilation operator of one vibrational quantum on molecule $n$.  The linear exciton--phonon coupling only is manifested in the treated crystals \cite{hoffmann02} and we get the following exciton--phonon part \cite{henessy99,hoffmann00,hoffmann02,lalov05}
\begin{equation}
\label{eq:exc-phon}
    \hat{H}_{\mbox{ex--phon}} = \sum_{n,\sigma=1,2}\left[ \xi_{\rm F}B_n^{+} B_n + \xi C_{n\sigma}^{+} C_{n\sigma} \right]\hbar \omega_0 \left( a_n^{+} + a_n \right),
\end{equation}
where $\xi_{\rm F}$ and $\xi$ are dimensionless parameters of the linear FE--phonon and CTEs--phonon coupling, correspondingly.

The full Hamiltonian $\hat{H}$ contains all the parts (\ref{eq:fehamilt})--(\ref{eq:exc-phon}) and can be transformed using the canonical transformation which eliminates the linear exciton--phonon coupling (\ref{eq:exc-phon}), see \cite{lalov05,davydov71},
\begin{equation}
\label{eq:h1}
    \hat{H}_1 = \exp(Q) \hat{H} \exp(-Q),
\end{equation}
where
\begin{equation}
\label{eq:q}
    Q = \sum_{n,\sigma=1,2} \left[ \xi_{\rm F}B_n^{+} B_n + \xi C_{n\sigma}^{+} C_{n\sigma} \right]\left( a_n^{+} - a_n \right).
\end{equation}

We introduce the vibronic operators
\begin{equation}
\label{eq:vn}
    V_n = \exp(Q) B_n \exp(-Q),
\end{equation}
\begin{equation}
\label{eq:unsigma}
    U_{n,\sigma} = \exp(Q) C_{n\sigma} \exp(-Q)
\end{equation}
and get the following transformed Hamiltonian
\begin{eqnarray}
\label{eq:h1new}
    \hat{H}_1 = \sum_n \left( E_{\rm F} - \hbar \omega_0 \xi_{\rm F}^2 \right)V_n^{+}V_n + \sum_{n n^{\prime}} L \left( \delta_{n^{\prime},n+1} + \delta_{n^{\prime},n-1} \right) V_{n^{\prime}}^{+} V_n \nonumber \\
    {}+ \sum_{n,\sigma} \left( E_{\rm c} - \hbar \omega_0 \xi^2 \right)U_{n \sigma}^{+}U_{n \sigma} + \sum_n \hbar \omega_0 a_n^{+} a_n \nonumber \\
    {}+ \sum_n \left[ \varepsilon_{\rm e1}V_n^{+}U_{n1} + \varepsilon_{\rm e2}V_n^{+}U_{n2} + \varepsilon_{\rm h1}V_n^{+}U_{n-1,1} + \varepsilon_{\rm h2}V_n^{+}U_{n+1,2} + \mbox{h.c.} \right]
\end{eqnarray}

In a stack with inversion center, point group $C_i$, the excitons are \emph{gerade\/} or \emph{ungerade\/} (in the center of the Brillouin zone, at $k = 0$).  The ungarade excitons only are dipole-active and influence the linear optical susceptibility and the absorption spectra.  In the case under consideration, the operator of the transition dipole moment has the following form \cite{lalov08,haarer75}:
\begin{equation}
\label{eq:p}
    P = \sum_n \left[ \mathbf{p}_{\rm F}\left( V_n^{+} + V_n \right) + \mathbf{p}_{\rm CT}\left( U_{n2} - U_{n1} + U_{n2}^{+} - U_{n1}^{+} \right)  \right].
\end{equation}

The gerade FEs, as well as the symmetrical combination of CTEs, $\left( U_{n2}^{+} + U_{n1}^{+} \right)|0\rangle$, can be also mixed, but due to their vanishing transition dipole moment they will not be considered here.

Introducing the Fourier transform in the momentum space of the vibronic operators,
\begin{equation}
\label{eq:vk}
    V_k = \frac{1}{\sqrt{N}}\sum_n V_n \exp(\mathrm{i}kna),
\end{equation}
\begin{equation}
\label{eq:uksigma}
    U_{k,\sigma} = \frac{1}{\sqrt{N}}\sum_n U_{n,\sigma} \exp(\mathrm{i}kna),
\end{equation}
we obtain the following form of the Hamiltonian (\ref{eq:h1new})
\begin{eqnarray}
\label{eq:h1new1}
    \hat{H}_1 = \sum_k \left( E_{\rm F} - \hbar \omega_0 \xi_{\rm F}^2 + 2L\cos k \right)V_k^{+}V_k + \sum_{k,\sigma} \left( E_{\rm c} - \hbar \omega_0 \xi^2 \right) U_{k\sigma}^{+}U_{k\sigma} \nonumber \\
    {}+ \sum_k \left\{ \left[ \varepsilon_{\rm e}^{\prime} + \varepsilon_{\rm h}^{\prime}\cos k + \mathrm{i}\varepsilon_{\rm h}^{\prime \prime}\sin k \right] V_k^{+}\left( U_{k,1} + U_{k,2} \right) \right. \nonumber \\
    \left.
    {}+ \left[ \varepsilon_{\rm e}^{\prime \prime} + \varepsilon_{\rm h}^{\prime \prime}\cos k + \mathrm{i}\varepsilon_{\rm h}^{\prime}\sin k \right] V_k^{+}\left( U_{k,2} - U_{k,1} \right) + \mbox{h.c.} \right\}
    + \sum_n \hbar \omega_0 a_n^{+} a_n,
\end{eqnarray}
where
\begin{equation}
\label{eq:epsprime}
    \varepsilon_{\rm e}^{\prime} = \left( \varepsilon_{\rm e1} + \varepsilon_{\rm e2} \right)/2, \qquad \varepsilon_{\rm h}^{\prime} = \left( \varepsilon_{\rm h1} + \varepsilon_{\rm h2} \right)/2,
\end{equation}
\begin{equation}
\label{eq:epssec}
    \varepsilon_{\rm e}^{\prime \prime} = \left( \varepsilon_{\rm e2} - \varepsilon_{\rm e1} \right)/2, \qquad \varepsilon_{\rm h}^{\prime \prime} = \left( \varepsilon_{\rm h2} - \varepsilon_{\rm h1} \right)/2.
\end{equation}
For the case of ungerade FEs, their mixing with the symmetrical (even) combination $\left( U_{k,1} + U_{k,2} \right)$ at $k = 0$ is impossible, and thus the mixing parameters $\varepsilon_{\rm e}^{\prime}$ and $\varepsilon_{\rm h}^{\prime}$ vanish, $\varepsilon_{\rm e}^{\prime} = \varepsilon_{\rm h}^{\prime} = 0$.  The final expressions for operators (\ref{eq:fcoperator}) and (\ref{eq:p}) are
\begin{eqnarray}
\label{eq:fc}
    \hat{H}_{\rm FC} = \sum_k \left[ \left( \varepsilon_{\rm e}^{\prime \prime} + \varepsilon_{\rm h}^{\prime \prime}\cos k \right) V_k^{+} \left( U_{k,2} - U_{k,1} \right) \right. \nonumber \\
    \left.
    {}+ \mathrm{i}\varepsilon_{\rm h}^{\prime \prime}\sin k V_k^{+} \left( U_{k,2} + U_{k,1} \right) + \mbox{h.c.} \right]
\end{eqnarray}
and
\begin{eqnarray}
\label{eq:pfinal}
    \hat{P} = \sqrt{N}\left[ \mathbf{p}_{\rm F}\left( V_{k=0} + V_{k=0}^{+} \right) \right. \nonumber \\
    \left.
    {}+ \mathbf{p}_{\rm CT}\left( U_{k=0,2} - U_{k=0,1} + U_{k=0,2}^{+} - U_{k=0,1}^{+} \right) \right],
\end{eqnarray}
respectively.

\section{Calculation of the linear optical susceptibility}
\label{sec:optical}

The linear optical susceptibility can be calculated by using the formula \cite{agranovich83}
\begin{equation}
\label{eq:chiij}
    \chi_{ij} = \lim_{\epsilon \to 0}\left\{ \frac{1}{2\hbar V}\left[ \Phi_{ij}(\omega + \mathrm{i}\epsilon) + \Phi_{ij}(-\omega + \mathrm{i}\epsilon) \right] \right\}
\end{equation}
with
\begin{equation}
\label{eq:phiij}
    \Phi_{ij}(t) = -\mathrm{i} \theta(t)\langle 0| \hat{P}_i(t)\hat{P}_j(0) + \hat{P}_j(t)\hat{P}_i(0)|0\rangle,
\end{equation}
where $V$ is the crystal's volume [in our case being proportional to $Nv$ ($v$ is the volume occupied by one molecule)] and $\hat{P}$ is the operator (\ref{eq:pfinal}).  The Green functions (\ref{eq:phiij}) have been calculated as average over the ground state $|0\rangle$ only by taking into account the large values of $E_{\rm F}$, $E_{\rm c}$, $\hbar \omega_0 \gg k_{\rm B}T$.

We calculate the Green functions (\ref{eq:phiij}) following the vibronic approach \cite{lalov07b,lalov08}.  In the next expression, the $x$ axis is supposed to be oriented along the vector $\mathbf{p}_{\rm F}$ which includes angle $\gamma$ with the vector $\mathbf{p}_{\rm CT}$, and $a = \left| p_{\rm CT}/p_{\rm F} \right|$.  Then we can represent the linear optical susceptibility of one stack as
\begin{equation}
\label{eq:chixx}
    \chi_{xx} = -\frac{p_{\rm F}^2}{v}\frac{1}{\alpha_1 \alpha_2 - 2\alpha_{12}^2} \left[ \alpha_2 + 4a\alpha_{12}\cos \gamma + 2a^2 \alpha_1 \cos^2 \gamma \right].
\end{equation}
The functions $\alpha_1$, $\alpha_2$, $\alpha_{12}$ have been calculated for the excitonic and one-phonon vibronic regions (see below).  The PTCDA and MePTCDI crystals contain two types $A$ and $B$ of parallel molecular stacks, however, the excitonic and vibronic excitations in each stack interact very weakly with the excitations of the other stacks.  In the same way as in Ref.~\cite{lalov06b}, we calculate the crystal's susceptibility in an oriented gas model.  We denote by $2\varphi$ the angle between the vectors $\mathbf{p}_{\rm F}^A$ and $\mathbf{p}_{\rm F}^B$ of two different stacks and suppose that these vectors are positioned in the $(XY)$ plane, the crystal $X$ axis been oriented along the sum $\mathbf{p}_{\rm F}^A + \mathbf{p}_{\rm F}^B$ \cite{note}.

The components of the linear optical susceptibility of the crystal correspondingly are:
\begin{eqnarray}
\label{eq:chixxnew}
    \chi_{XX} = -\frac{p_{\rm F}^2}{v}\frac{2}{\alpha_1 \alpha_2 - 2\alpha_{12}^2} \left[ \alpha_2 \cos^2 \varphi \right. \nonumber \\
    \left.
    {}+ 4a\alpha_{12}\cos^2 \varphi \cos \gamma + a^2 \alpha_1 (1 + \cos 2\gamma \cos \varphi) \right]
\end{eqnarray}
and
\begin{eqnarray}
\label{eq:chiyy}
    \chi_{YY} = -\frac{p_{\rm F}^2}{v}\frac{2}{\alpha_1 \alpha_2 - 2\alpha_{12}^2} \left[ \alpha_2 \sin^2 \varphi \right. \nonumber \\
    \left.
    {}+ 4a\alpha_{12}\sin^2 \varphi \cos \gamma + a^2 \alpha_1 (1 - \cos 2\gamma \cos \varphi) \right].
\end{eqnarray}

We find the following expressions for functions $\alpha_1$, $\alpha_2$, $\alpha_{12}$:

(1) In the excitonic region expressions practically coincide with formulas in Ref.~\cite{lalov08}, namely
\begin{equation}
\label{eq:alpha1}
    \alpha_1 = \hbar \omega - \left( E_{\rm F} + 2L \right) - \hbar \Omega_{\rm 0F}(1),
\end{equation}
\begin{equation}
\label{eq:alpha12}
    \alpha_{12} = \varepsilon_{\rm e}^{\prime \prime} + \varepsilon_{\rm h}^{\prime \prime},
\end{equation}
\begin{equation}
\label{eq:alpha2}
    \alpha_2 = \hbar \left[ \omega - \Omega_{\rm 0c}(1) \right] - E_{\rm c},
\end{equation}
where $\Omega_{\rm 0F}(1)$ and $\Omega_{\rm 0c}(1)$ are expressed through the continuous fractions following from recursions:
\begin{equation}
\label{eq:omega0f}
    \Omega_{\rm 0F}(n) = \frac{n\omega_{\rm a}^2}{\omega - \left( E_{\rm F} + 2L \right)/\hbar - \omega_0 - \Omega_{\rm 0F}(n+1)},
\end{equation}
\begin{equation}
\label{eq:omega0c}
    \Omega_{\rm 0c}(n) = \frac{n\omega_{\rm a1}^2}{\omega - E_{\rm c}/\hbar - \omega_0 - \Omega_{\rm 0c}(n+1)},
\end{equation}
\begin{equation}
\label{eq:omegaa}
    \omega_{\rm a} = \xi_{\rm F}\omega_0, \qquad \omega_{\rm a1} = \xi \omega_0.
\end{equation}

(2) In the one-phonon vibronic region
\begin{equation}
\label{eq:alfa1}
    \alpha_1 = \hbar \omega - \left( E_{\rm F} + 2L \right) - \frac{\omega_{\rm a}^2}{D} \left[ \sigma_2 + MT \right],
\end{equation}
\begin{equation}
\label{eq:alfa12}
    \alpha_{12} = \varepsilon_{\rm e}^{\prime \prime} + \varepsilon_{\rm h}^{\prime \prime} + \frac{\omega_{\rm a} \omega_{\rm a1}}{D}\sigma_3,
\end{equation}
\begin{equation}
\label{eq:alfa2}
    \alpha_2 = \hbar \omega - E_{\rm c} - \frac{\omega_{\rm a1}^2}{D} \left[ \sigma_4 + M_{\rm F}T \right],
\end{equation}
where
\begin{equation}
\label{eq:sigma2}
    \sigma_2 = \frac{\hbar \left[ \omega - \omega_0 - \Omega_{\rm 1c}(1) \right]- E_{\rm c}}{m}\sigma,
\end{equation}
\begin{equation}
\label{eq:sigma3}
    \sigma_3 = (1/m)\left( \varepsilon_{\rm e}^{\prime \prime}\sigma + \varepsilon_{\rm h}^{\prime \prime}\sigma_1 \right),
\end{equation}
\begin{equation}
\label{eq:sigma4}
    \sigma_4 = \frac{1}{\hbar \left[ \omega - \omega_0 - \Omega_{\rm 1c}(1) \right]- E_{\rm c}}\left\{ 1 + \frac{2}{m}
    \left[ \left( \varepsilon_{\rm e}^{\prime \prime} \right)^2 \sigma + \varepsilon_{\rm h}^{\prime \prime} \sigma_1 \left( 2\varepsilon_{\rm e}^{\prime \prime} + \varepsilon_{\rm h}^{\prime \prime}t \right) \right] \right\},
\end{equation}
\vspace{0.5mm}
\begin{equation}
\label{eq:tcap}
    T = -\left[ \frac{\sigma}{m} + \frac{2 \left(\varepsilon_{\rm h}^{\prime \prime}\right)^2 \sigma_1}{m^2} \right],
\end{equation}
\vspace{0.5mm}
\begin{equation}
\label{eq:m}
    m = 2\left\{ L\left[ \hbar \left[ \omega - \omega_0 - \Omega_{\rm 1c}(1) \right]- E_{\rm c} \right] + 2\varepsilon_{\rm e}^{\prime \prime} \varepsilon_{\rm h}^{\prime \prime} \right\},
\end{equation}
\vspace{-4mm}
\begin{eqnarray}
\label{eq:t}
    t = (1/m)\left\{ \phantom{\left[\left( \varepsilon_{\rm e}^{\prime \prime} \right)^2\right]}\!\!\!\!\!\!\!\!\!\!\!\!\!\!\!\!\!\!
    \left\{ \hbar \left[ \omega - \omega_0 - \Omega_{\rm 1F}(1) \right]- E_{\rm F} \right\} \left\{ \hbar \left[ \omega - \omega_0 - \Omega_{\rm 1c}(1) \right]- E_{\rm c} \right\} \right. \nonumber \\
    \left.
    {}- 2\left[ \left( \varepsilon_{\rm e}^{\prime \prime} \right)^2 + \left( \varepsilon_{\rm h}^{\prime \prime} \right)^2 \right] \right\},
\end{eqnarray}
\begin{equation}
\label{eq:d}
    D = 1 - M_{\rm F}\sigma_2 - M\sigma_4 - M_{\rm F}MT,
\end{equation}
\begin{equation}
\label{eq:sigma}
    \sigma = \left\{ \begin{array}{cc}
    -1/\sqrt{t^2 - 1} & \mbox{if\,\,\, $\left| t + \sqrt{t^2 - 1} \right| < 1$,}
    \\ \\
    1/\sqrt{t^2 - 1} & \mbox{if\,\,\, $\left| t + \sqrt{t^2 - 1} \right| < 1$,}
                     \end{array}
    \right.
\end{equation}
and
\begin{equation}
\label{eq:sigma1}
    \sigma_1 = t\sigma -1.
\end{equation}

The functions $\Omega_{\rm 1F}(1)$ and $\Omega_{\rm 1c}(1)$ also represent continuous fractions from recursions:
\begin{equation}
\label{eq:omega1f}
    \Omega_{\rm 1F}(n) = \frac{n\omega_{\rm a}^2}{\omega - E_{\rm F}/\hbar - (n+1)\omega_0 - \Omega_{\rm 1F}(n+1)},
\end{equation}
\begin{equation}
\label{eq:omega1c}
    \Omega_{\rm 1c}(n) = \frac{n\omega_{\rm a}^2}{\omega - E_{\rm c}/\hbar - (n+1)\omega_0 - \Omega_{\rm 1c}(n+1)}.
\end{equation}
Finally, the functions $M$ and $M_{\rm F}$ can be expressed as follows:
\begin{equation}
\label{eq:m/h}
    M/\hbar = \Omega_{\rm 0c}(2) - \Omega_{\rm 1c}(1),
\end{equation}
\begin{equation}
\label{eq:mf/h}
    M_{\rm F}/\hbar = \Omega_{\rm 0F}(2) - \Omega_{\rm 1F}(1).
\end{equation}

\section{Simulations of the excitonic and vibronic spectra of MePTCDI and PTCDA crystals}
\label{sec:simul}
In this section, we calculate the absorption spectra of the two crystals finding the imaginary parts of the components of the linear optical susceptibility, (\ref{eq:chixxnew}) and (\ref{eq:chiyy}), at $\left( 2p_{\rm F}^2/v \right) = 1$ and supposing an imaginary part equal to $\mathrm{i}\delta/\hbar$ of the frequency $\omega$.  We put the excitonic and vibrational parameters for the studied crystals as they have been fitted in Refs.~\cite{henessy99} and \cite{schmidt02} and used in our previous papers \cite{lalov06a,lalov07a,lalov06b,lalov06c}, see table 1:
\begin{table}
\caption{\label{data}Excitonic and vibrational parameters (in cm$^{-1}$) of the MePTCDI and PTCDA crystals.}\vspace*{-3mm}
\begin{center}
\hspace*{25mm}\begin{tabular}{lccccrccc}
\br
 & $E_{\rm F}$ & $E_{\rm c}$ & $\hbar \omega_0$ & $L$ & $\varepsilon_{\rm e}$ & $\varepsilon_{\rm h}$ & $2\varphi$ & $\gamma$ \\
\mr
MePTCDI & $17\,992$ & $17\,346$ & $1\,400$ & $355$ & $-380$ & $-137$ & $36.8^\circ$ & $68.1^\circ$ \\
PTCDA   & $18\,860$ & $18\,300$ & $1\,400$ & $330$ & $-48$ & $-436$ & $82^\circ$ & $143^\circ$ \\
\br
\end{tabular}
\end{center}
\end{table}
The data for angle $\gamma$ and ratio $a = \left| p_{\rm CT}/p_{\rm F} \right|$ have been calculated in \cite{hoffmann00} using quantum chemical evaluations.  We use the values $a = 0.1$ for the PTCDA crystal and $a = 0.135$ for the MePTCDI.  The data for angle $2\varphi$ are derived from the crystal structure (for MePTCDI see \cite{haedicke86}).

The aforementioned parameters are permanent in our calculations.  We vary the values of the following parameters (intending to observe their impact on the absorption spectra and find a better similarity with the experimental absorption spectra \cite{henessy99,hoffmann00,hoffmann02}):

(i) The excitonic damping quantity $\delta$ which varies from $80$ cm$^{-1}$ to $500$ cm$^{-1}$.

(ii) The linear exciton--phonon coupling parameters $\xi_{\rm F}$ and $\xi$ which vary near the values
\[
    \xi_{\rm F} = 0.82, \quad \xi = \xi_{\rm F}/\sqrt{2} \quad \mbox{for PTCDA (see \cite{henessy99})},
\]
\[
    \xi_{\rm F} = 0.88, \quad \xi = \xi_{\rm F}/\sqrt{2} \quad \mbox{for MePTCDI (see \cite{hoffmann02})}.
\]
In calculating continuous fractions (\ref{eq:omega0f}), (\ref{eq:omega0c}) and (\ref{eq:omega1f}), (\ref{eq:omega1c}) we take twenty steps.

\subsection{MePTCDI}
\label{subsec:meptcdi}
\begin{figure}[!ht]
\centering\includegraphics[height=.30\textheight]{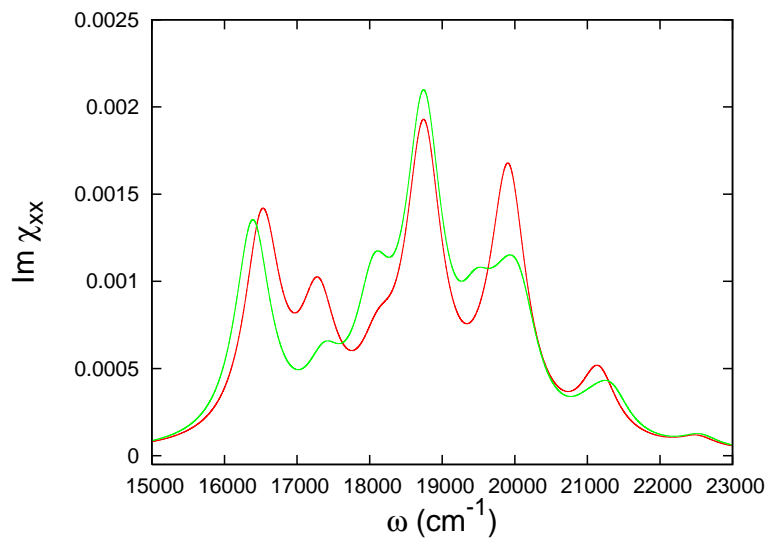}
  \caption{Linear absorption of the MePTCDI crystal at $\delta = 300$ cm$^{-1}$ and $\xi_{\rm F} = 0.88$.  Red curve is calculated for the excitonic region [using formulas (\ref{eq:alpha1})--(\ref{eq:alpha2})] and the green curve for one-phonon vibronic spectra [formulas (\ref{eq:alfa1})--(\ref{eq:alfa2})].}
  \label{fig:fg1}
\end{figure}

The general structure of the absorption spectra at $\delta = 300$ cm$^{-1}$ can be seen in figure~\ref{fig:fg1} in which the red curve is calculated using `excitonic' formulas (\ref{eq:alpha1})--(\ref{eq:alpha2}) (in the following denoted as `exc' program) and the green curve by using one-phonon vibronic formulas (\ref{eq:alfa1})--(\ref{eq:alfa2}) (the corresponding program is denoted as `1p').  The approximate boundary between the two regions is $17\,800$ cm$^{-1}$.  Two absorption maxima associated with the two types of excitons (FEs and CTEs) appear in the excitonic spectra below $17\,800$ cm$^{-1}$, as well as in the vibronic spectra (green curve).  Our calculations are valid for the one-phonon vibronic spectrum, approximately between $17\,800$ cm$^{-1}$ and $19\,100$ cm$^{-1}$ but the higher vibronics also appear in the calculated spectra and they show splitting associated with a FE--CTEs mixing (see the doublet near $19\,500$--$20\,000$ cm$^{-1}$).  The shape of the experimental absorption curve, see figures 2 and 3 from Ref.~\cite{hoffmann00}, can be combined approximately using the excitonic curve (up to $17\,800$ cm$^{-1}$) and the green curve above that frequency.

\begin{figure}[!ht]
\centering\includegraphics[height=.30\textheight]{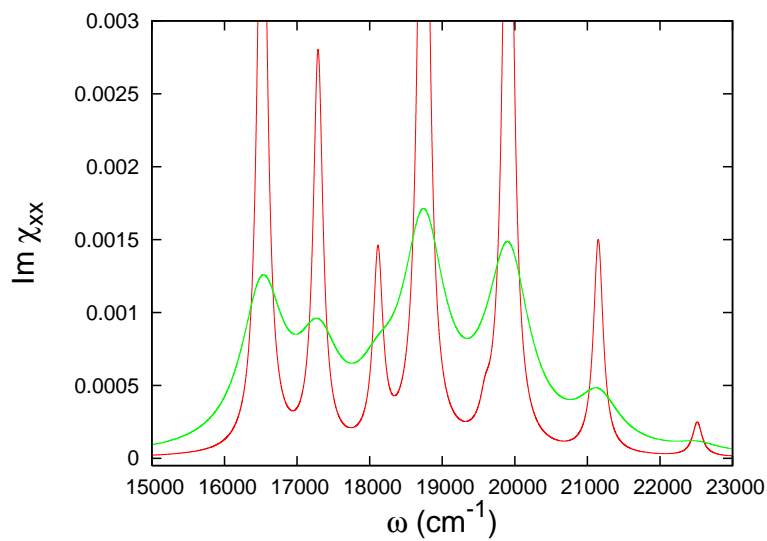}
  \caption{Absorption spectra calculated with program `exc' for the following values of the excitonic damping: red curve with $\delta = 80$ cm$^{-1}$ and the green one with $\delta = 350$ cm$^{-1}$.}
  \label{fig:fg2}
\end{figure}

The impact of the FE--CTEs mixing can be seen more clearly by comparing the two curves in figure~\ref{fig:fg2} calculated with the `exc' program for two different values of the damping $\delta$.  The smaller value $\delta = 80$ cm$^{-1}$ generates sharp absorption maxima in the excitonic and one-phonon vibronic regions.

\begin{figure}[!ht]
\centering\includegraphics[height=.30\textheight]{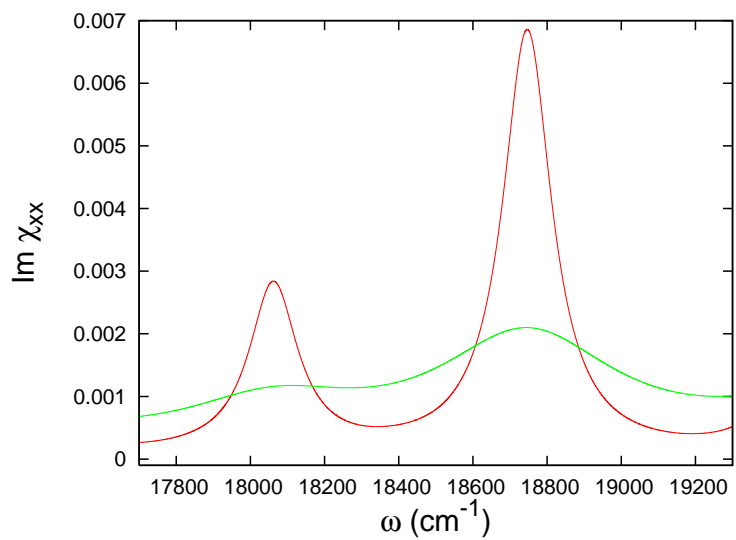}
  \caption{One-phonon vibronic absorption spectra at $\xi_{\rm F} = 0.88$, $\delta = 80$ cm$^{-1}$ (red) and  $\delta = 350$ cm$^{-1}$ (green).}
  \label{fig:fg3}
\end{figure}

The one-phonon vibronic spectrum calculated with the `1p' program is presented in figure~\ref{fig:fg3}.  The maximum near $18\,800$ cm$^{-1}$ at $\delta = 300$ cm$^{-1}$ would be relatively flat but still resolvable.

\begin{figure}[!ht]
\centering\includegraphics[height=.30\textheight]{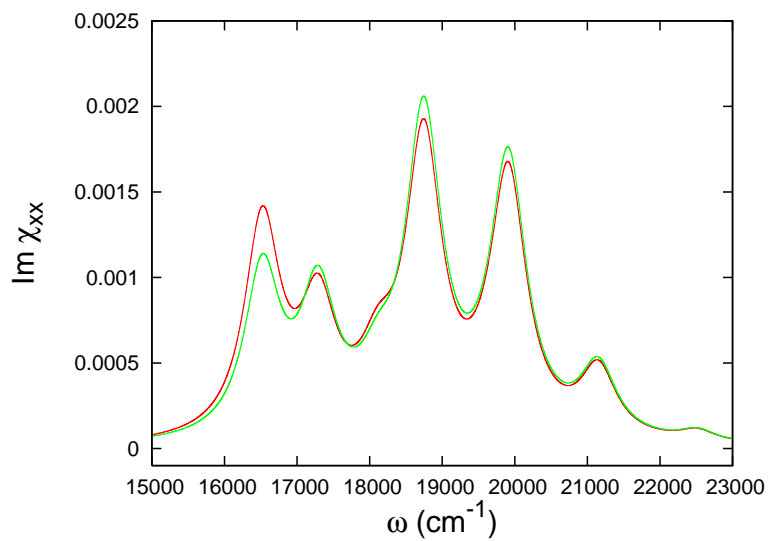}
  \caption{Impact of the CTEs-transition dipole moment expressed through the ratio $a$.  Red curve for $a = 0.135$ and the green one for $a = 0$.}
  \label{fig:fg4}
\end{figure}

The impact of the CTEs transition dipole moment is illustrated in figure~\ref{fig:fg4}.  The two curves---the red calculated with the CTEs contribution, and the green one calculated without this contribution [$a \equiv 0$ in formulas (\ref{eq:chixx})--(\ref{eq:chiyy})]---are relatively close to each other.  The strongest impact can be measured near lower excitonic peak generated by the CTEs level $E_{\rm c} = 17\,346$ cm$^{-1}$.

\begin{figure}[!ht]
\centering\includegraphics[height=.30\textheight]{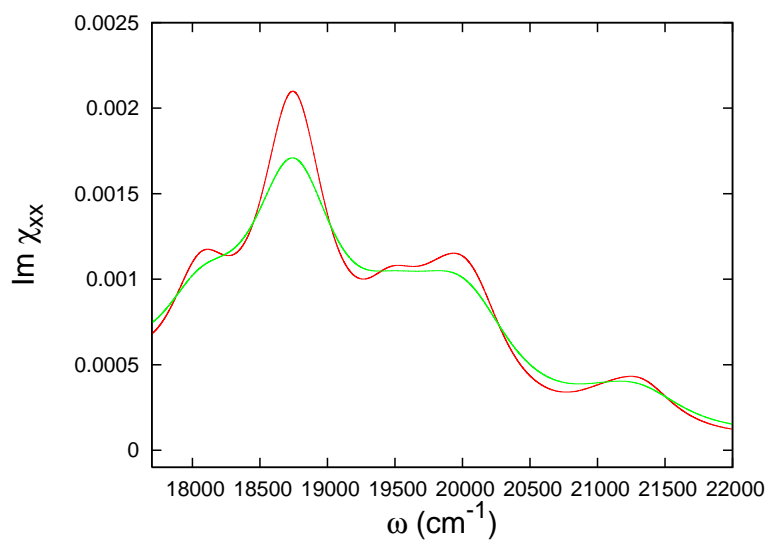}
  \caption{Vibronic absorption at various values of $\delta$: red curve for $\delta = 300$ cm$^{-1}$ and the green curve for $\delta = 400$ cm$^{-1}$.}
  \label{fig:fg5}
\end{figure}

Figure~\ref{fig:fg5} demonstrates the changes of the absorption lineshape in the vibronic spectra depending on the values of the damping $\delta$.  The bigger value of $\delta$ (the green curve) makes the absorption curve flat.  The half-width of the highest red maximum approximately coincides with the magnitude of $\delta$ ($= 300$ cm$^{-1}$) which confirms its one-particle nature (bound exciton--phonon state).

\begin{figure}[!ht]
\centering\includegraphics[height=.30\textheight]{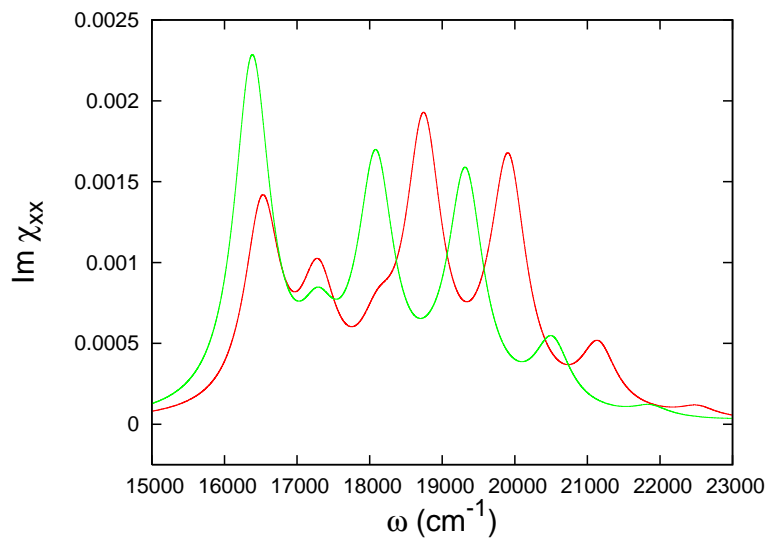}
  \caption{Comparison of absorption curves calculated with data from table 1 [red and green curves calculated for $E_{\rm F} = 17\,346$ cm$^{-1}$ and $E_{\rm c} = 17\,992$ cm$^{-1}$, respectively (interchange of the excitonic levels)].}
  \label{fig:fg6}
\end{figure}

The two curve in figure~\ref{fig:fg6} calculating by using the `exc' program for $\delta = 300$ cm$^{-1}$ differ by the mutual positions of the two excitonic levels. The red curve corresponds to a lower position of the CTEs level $E_{\rm c} < E_{\rm F}$, whereas the green curve corresponds to the opposite situation $E_{\rm F} < E_{\rm c}$ (the magnitudes of $E_{\rm c}$ and $E_{\rm F}$ are exchanged).  The green curve exhibits a strong domination of the lower maxima associated with the FE level. The two curves can be approximated with five Lorentz maxima in the spectral region of $15\,000$--$22\,000$ cm$^{-1}$ \cite{hoffmann00}.  Comparing the lineshapes of these two curves with the experimental curve in Ref.~\cite{hoffmann00}, we cannot make a hypothesis which possibility is more probable.  We prefer the fitting from Refs.~\cite{hoffmann00} and \cite{schmidt02} $\left( E_{\rm c} < E_{\rm F} \right)$, however, another choice of the mixing arrangements is also allowed.

\begin{figure}[ht]
  \begin{minipage}[b]{0.5\linewidth}
  \centering
    \includegraphics[width=7.5cm]{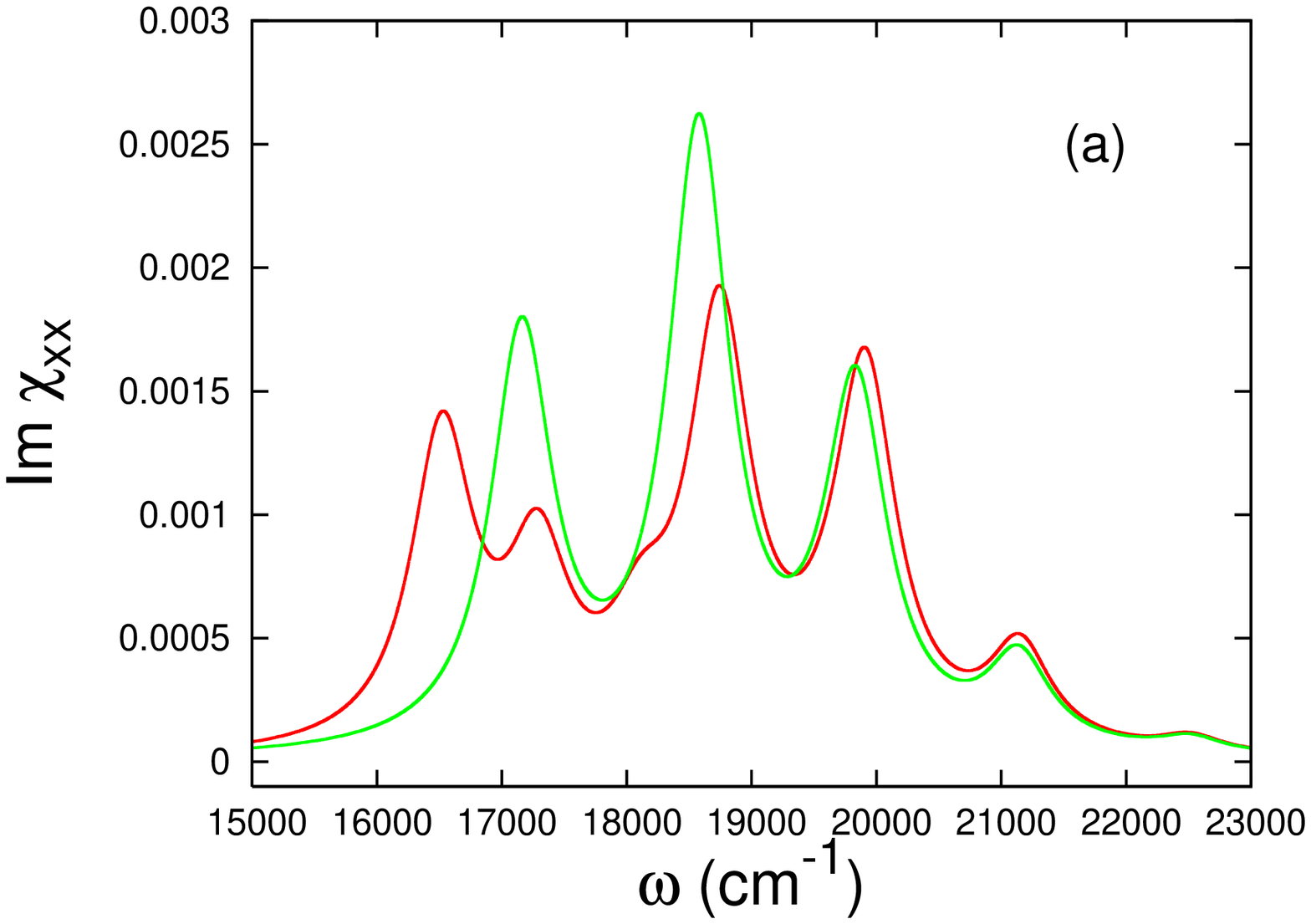}
  \end{minipage}
  \hspace{-0.2cm}
  \begin{minipage}[b]{0.5\linewidth}
  \centering
    \includegraphics[width=7.5cm]{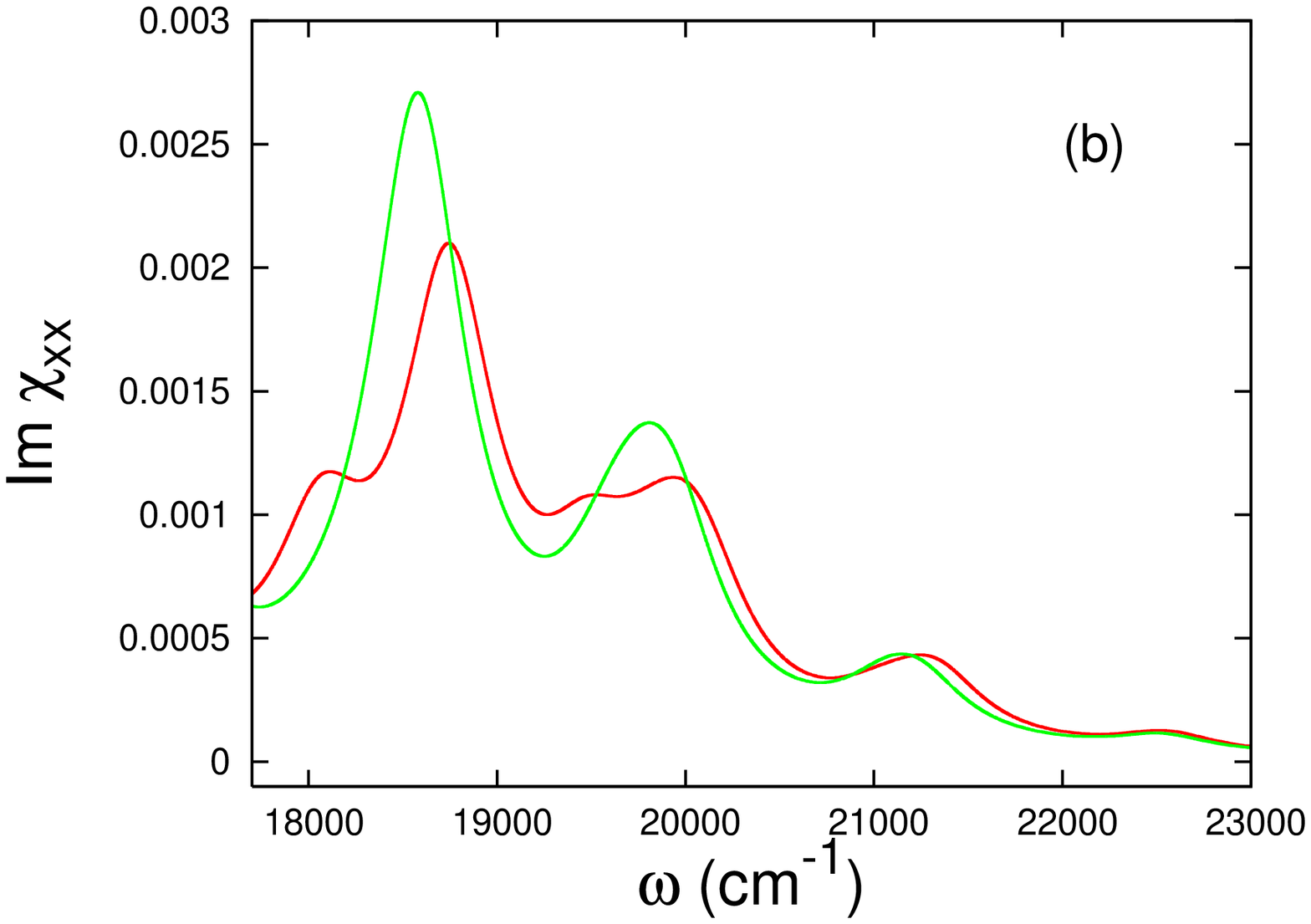}
  \end{minipage}
  \caption{Linear absorption spectra with FE--CTEs mixing (red curves) and pure Frenkel exciton spectra (green curves at $\varepsilon_{\rm e}^{\prime \prime} = \varepsilon_{\rm h}^{\prime \prime} = 0$), $\delta = 300$ cm$^{-1}$: (a) absorption spectra calculated with `exc' and (b) `1p' absorption curves.}
  \label{fig:fg7}
\end{figure}

The importance of the FE--CTEs mixing can be seen in figure~\ref{fig:fg7} in which the green curves represent a pure FE absorption $\left( \varepsilon_{\rm e}^{\prime \prime} = \varepsilon_{\rm h}^{\prime \prime} = 0 \right)$.  The complexity of the experimental absorption curves with five Lorentz maxima in the studied spectral region, see figures 2 and 3 in Ref.~\cite{hoffmann00}, cannot be understood on the basis of simple Frenkel-exciton model.  Contrary, the model of mixed FE--CTEs and their vibronic satellites can be the basis of adequate simulations of the absorption curves (red curves).

\begin{figure}[!ht]
\centering\includegraphics[height=.30\textheight]{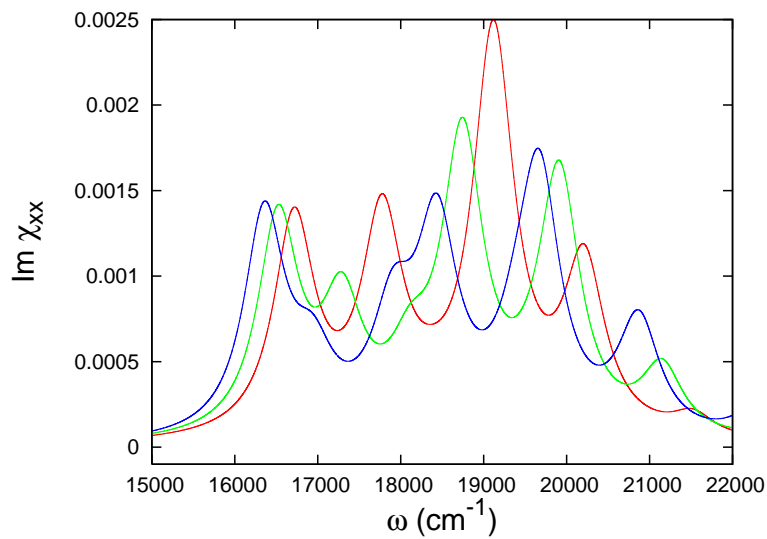}
  \caption{Linear absorption spectra `exc' at $\delta = 300$ cm$^{-1}$ calculated for the following values of the FE--phonon coupling constant $\xi_{\rm F}$: red curve for $\xi_{\rm F} = 0.7$, green curve for $\xi_{\rm F} = 0.88$, and blue curve for $\xi_{\rm F} = 1$.}
  \label{fig:fg8}
\end{figure}

Figure~\ref{fig:fg8} contains the calculated absorption curves at $\delta = 300$ cm$^{-1}$ (with the `exc' program) for three different values of the linear exciton--phonon coupling constant $\xi_{\rm F}$.  The values of $\xi_{\rm F}$ which we put in calculations can be considered as a characteristic of the intermediate coupling, but the lineshape of the absorption curves is quite different.  Obviously only the green curve exhibits five absorption maxima despite of the four maxima and the two saddle parts of the blue curve.  The value $\xi_{\rm F} = 0.88$ which we use in calculating other cases (figures~\ref{fig:fg1}--\ref{fig:fg7}) seems to be the most probable.

The model of the FE--CTEs mixing reproduces the general structure of the absorption spectra in the excitonic and vibronic regions of the MePTCDI crystal.  Our calculation confirm the correct choice of the excitonic and vibrational parameters in the fitting procedure implemented in \cite{hoffmann00,schmidt02}.  We establish as the most probable value of the exciton damping $\delta = 300$ cm$^{-1}$.  For us, however, one open question still exists, notably the mutual positions of the excitonic levels of FE and CTEs lines.

\subsection{PTCDA}
\label{subsec:ptcda}
The experimental linear absorption lines for the PTCDA crystal are broader than those of the MePTCDI (see \cite{hoffmann00}).  Thus the structure of the PTCDA spectra does not exhibit many details due to the FE--CTEs mixing.

\begin{figure}[!ht]
\centering\includegraphics[height=.30\textheight]{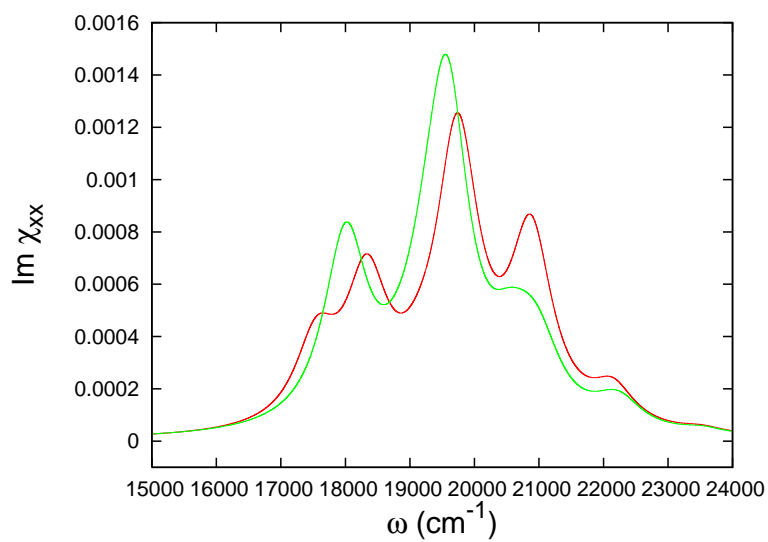}
  \caption{Linear absorption of the PTCDA crystal at $\xi_{\rm F} = 0.82$ and $\delta = 400$ cm$^{-1}$.  Red curve for the `exc' calculations and the green one for the `1p' calculations.}
  \label{fig:fg9}
\end{figure}

In figure~\ref{fig:fg9} the boundary between the excitonic and vibronic spectra is around $18\,800$ cm$^{-1}$ and the FE--CTEs splitting can be observed in the red `exc' curve only.  The splitting in the vibronic satellites is practically unresolvable.

\begin{figure}[!ht]
\centering\includegraphics[height=.30\textheight]{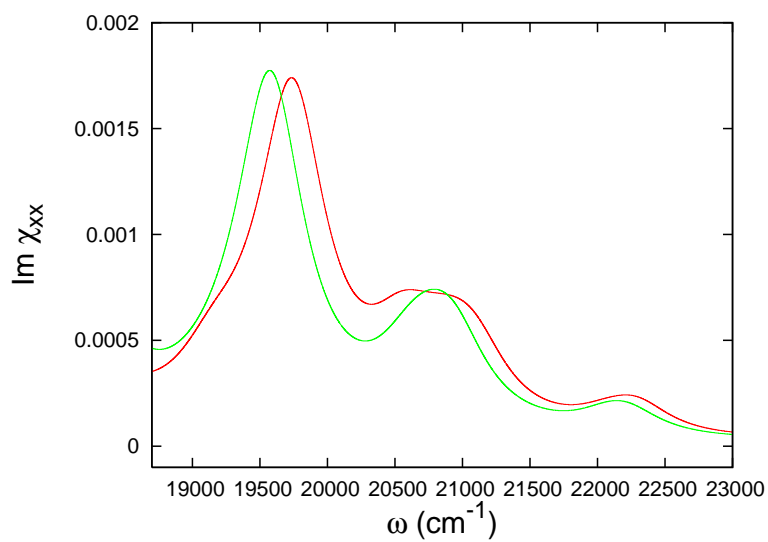}
  \caption{Linear absorption of PTCDA (`1p' program) at $\xi_{\rm F} = 0.82$ and $\delta = 400$ cm$^{-1}$ with FE--CTEs mixing (red curve) and without it (green curve with $\varepsilon_{\rm e}^{\prime \prime} = \varepsilon_{\rm h}^{\prime \prime} = 0$).}
  \label{fig:fg10}
\end{figure}

Figure~\ref{fig:fg10} demonstrates relatively a weak impact of the FE--CTEs mixing on the vibronic spectra where the pure FE satellites (green colour) and the satellites of the mixed excitons (red colour) lie very close.

\begin{figure}[!ht]
\centering\includegraphics[height=.30\textheight]{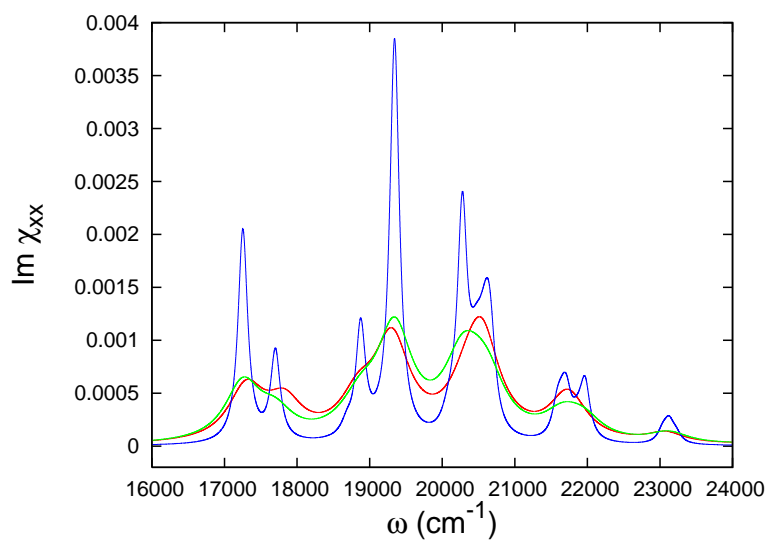}
  \caption{Linear absorption of PTCDA at $\xi_{\rm F} = 1$.  Red curve (`exc') at $\delta = 300$ cm$^{-1}$, green curve (`1p') at $\delta = 300$ cm$^{-1}$, and blue curve (`1p') at $\delta = 80$ cm$^{-1}$.}
  \label{fig:fg11}
\end{figure}

The absorption curves in figure~\ref{fig:fg11} have been calculated with $\xi_{\rm F} = 1$.  The excitonic and vibronic lines at $\delta = 300$ cm$^{-1}$ demonstrate doublet-like structure confirmed on the blue curve corresponding to a non-realistic low damping $\delta = 80$ cm$^{-1}$.  The red and green curves are similar to the lineshape of the experimental curves.

\begin{figure}[!ht]
\centering\includegraphics[height=.30\textheight]{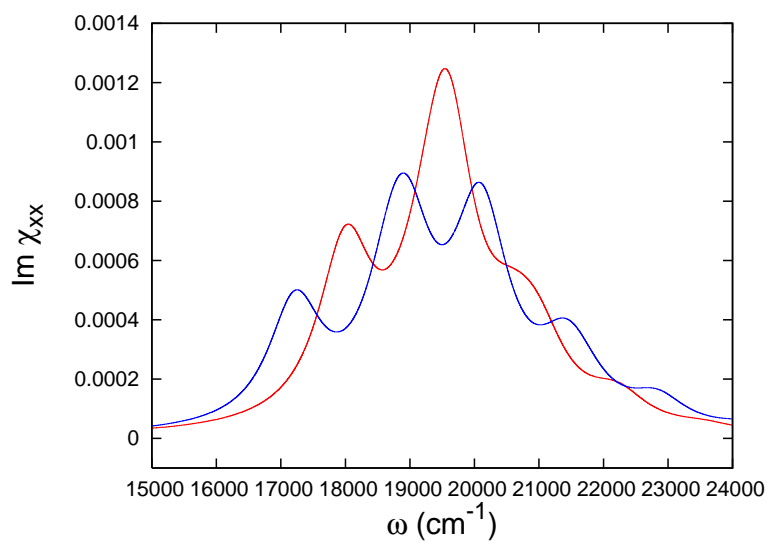}
  \caption{Linear absorption of PTCDA at $\delta = 500$ cm$^{-1}$.  Red curve (`1p') corresponds to $\xi_{\rm F} = 0.82$ and the blue curve (`1p')---to $\xi_{\rm F} = 1.1$.}
  \label{fig:fg12}
\end{figure}

\begin{figure}[!ht]
\centering\includegraphics[height=.30\textheight]{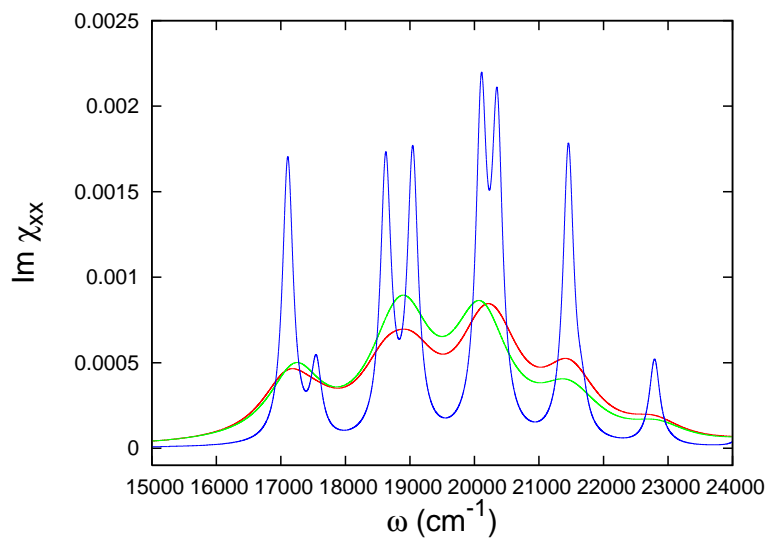}
  \caption{Linear absorption of PTCDA at $\xi_{\rm F} = 1.1$.  Red curve (`exc') and green curve (`1p') correspond to $\delta = 500$ cm$^{-1}$, while the blue curve (`exc')---to $\delta = 100$ cm$^{-1}$.}
  \label{fig:fg13}
\end{figure}

The real width of the excitonic curve is about $\delta = 500$ cm$^{-1}$ (figure~\ref{fig:fg12}).  The absorption above $18\,500$ cm$^{-1}$ with $\xi_{\rm F} = 0.82$ (red curve) can be described with one maximum near $19\,500$ cm$^{-1}$ and its higher vibronic replicas.  The blue curve calculated with $\xi_{\rm F} = 1.1$ is more similar to the experimental curve (see \cite{henessy99}) whose lineshape covers several vibronic maxima.  We accept as a very good simulation of the absorption spectra namely the case of $\delta = 500$ cm$^{-1}$, $\xi_{\rm F} = 1.1$.  It can be seen in figure~\ref{fig:fg13} that the excitonic and first two vibronic maxima represent doublets whose splitting results from the FE--CTEs mixing, but this doublet structure is hidden due to the big damping.  The blue curve calculated at $\delta = 100$ cm$^{-1}$ shows the true structure of the wide absorption maxima.

\begin{figure}[!ht]
\centering\includegraphics[height=.30\textheight]{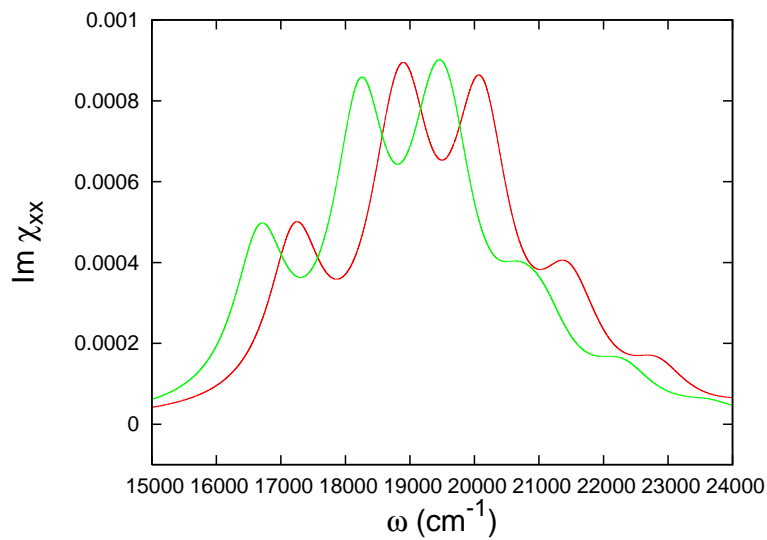}
  \caption{Linear absorption of PTCDA at $\xi_{\rm F} = 1.1$ and $\delta = 500$ cm$^{-1}$ (`1p' program).  The two excitonic levels are interchanged and the red curve corresponds to the values in table 1 ($E_{\rm c} < E_{\rm F}$) whereas for the green curve $E_{\rm F} < E_{\rm c}$.}
  \label{fig:fg14}
\end{figure}

Figure~\ref{fig:fg14} shows insensitivity of the absorption lineshape in the case of a mutual replacement of the two excitonic levels $E_{\rm c} < E_{\rm F}$ and $E_{\rm F} < E_{\rm c}$.  Obviously the absorption curves are shifted, however, their lineshapes are very similar.

So, in the PTCDA crystal the large excitonic linewidth which we estimate to be near $\delta = 500$ cm$^{-1}$ covers the effects of the FE--CTEs mixing.  Another result of the simulations of the absorption spectra is the hypothesis for a higher value of the constant $\xi_{\rm F}$ of the linear exciton--phonon coupling evaluated up to now as $\xi_{\rm F} = 0.82$ \cite{henessy99,hoffmann02}.  Our calculations show that possible values of $\xi_{\rm F}$ should be $1$ or $1.1$.

\section{Conclusions}
\label{sec:concl}
Our model for the linear absorption spectra of the molecular crystals includes the following parameters:
\begin{itemize}
\item excitonic levels $E_{\rm F}$ and $E_{\rm c}$, as well as the vibrational frequency $\omega_0$ of the intramolecular mode,
\item parameters $\varepsilon_{\rm e}^{\prime \prime}$ and $\varepsilon_{\rm h}^{\prime \prime}$ of the FE--CTEs mixing,
\item constants $\xi_{\rm F}$ and $\xi$ of the linear exciton--phonon coupling,
\item the width $\delta$ of the excitonic linewidth,
\item angles $2\varphi$, $\gamma$ and the ratio $a$ of the CTEs and FE transition dipole moments.
\end{itemize}

Practically all parameters have been introduced and fitted for the MePTCDI and PTCDA crystals in previous papers \cite{henessy99,hoffmann00,hoffmann02,schmidt02}.  In our study we apply the complex vibronic approach in calculating the linear optical susceptibility and its imaginary part which is a factor in the absorption coefficient in the excitonic and one-phonon vibronic spectra.  Higher vibronics---with two, three phonons---also have been demonstrated in our calculations.  The main goal of our model is to simulate the lineshape in the absorption region of $15\,000$--$23\,000$ cm$^{-1}$ of the aforementioned crystals, and to find out the adequate values of the excitonic linewidth, exciton--phonon coupling parameters and so on.  The better coincidence between the calculated and measured absorption curves can be achieved by fitting all the parameters of the model, especially the positions and distances between the excitonic levels.  The inclusion of external phonons can also cause some changes in the simulations \cite{brillante80}, as well as the inclusion of several intramolecular vibrational modes.

We stress again the main conclusions which concern the two crystals:
\begin{enumerate}
\item The excitonic linewidth of the MePTCDI crystal, according to our simulations, is approximately $\delta \approx 300$ cm$^{-1}$.  The FE-phonon linear coupling coefficient $\xi_{\rm F}$ has been estimated correctly in the previous papers \cite{henessy99,hoffmann02} as $\xi_{\rm F} \approx 0.88$.  The absorption spectra depend strongly on the mutual position of the two excitonic levels $E_{\rm F}$ and $E_{\rm c}$, but the simulations of the absorption spectra in our paper do not give the opportunity to choose one of the two possibilities: $E_{\rm F} < E_{\rm c}$ or $E_{\rm c} < E_{\rm F}$.
\item The excitonic linewidth of the PTCDA crystal can be evaluated as $\delta \approx 500$ cm$^{-1}$.  That is why the effect of the FE--CTEs mixing, being covered by the wide absorption maxima, are weakly expressed than in the MePTCDI crystal.  However, a very probable conclusion from our calculations may be the stronger linear exciton--phonon coupling ($\xi_{\rm F} \approx 1$ or $1.1$ instead of $0.82$).
\end{enumerate}

A supposition for FE--CTEs mixing is necessary for the interpretation of the excitonic and vibronic spectra of the MePTCDI crystal.  Concerning the PTCDA crystal our calculations suggest two possible models: (a) the FE--CTEs model studied in this paper, and (b) the model of pure FE and its vibronic satellites assuming wide excitonic levels.

Our model can be applied in the interpretation of other one-component molecular stacks (crystals) and it can be more effective in the systems with narrower excitonic absorption lines.

\clearpage
\large{\centerline{\textbf{Figures and Figure Captions}}}

\begin{thebibliography}{25}
\bibitem{note}              
In the crystals under consideration vectors $\mathbf{p}_{\rm CT}^{A}$ and $\mathbf{p}_{\rm CT}^{ B}$ are situated very close to the same plane ($XY$), see \cite{hoffmann00}.
\end{thebibliography}
\end{document}